\documentstyle[preprint,epsf,aps]{revtex}
\begin{document}
\title{Quantum error rejection code with spontaneous 
parametric down conversion}
\author{Xiang-Bin Wang\thanks{email: wang$@$qci.jst.go.jp}
\\
        Imai Quantum Computation and Information project, ERATO, Japan Sci. and Tech. Corp.\\
Daini Hongo White Bldg. 201, 5-28-3, Hongo, Bunkyo, Tokyo 113-0033, Japan}

\maketitle 
\begin{abstract}
We propose a linear optical scheme to transmit an unknown qubit
robustly over bit-flip-error channel. To avoid the technical difficulty
of the standard quantum error correction code, our scheme is based
on the concept of error-rejection. The whole scheme is  based on
currently existing technology. 
\end{abstract}
\section{Introductoin}
An unknown qubit can be sent to a remote party robustly 
through a noisy channel if we use the quantum error correction 
code (QECC)\cite{shor,steane,puri}, which plays a very important role in 
quantum computation and information\cite{chuang}.
The main idea there is first to encode the unknown qubit to an entangled
state of many qubits and after the remote party receives this 
quantum code, he  first decodes it and then 
obtains the  original state faithfully.
This is very different from the classical error correction since the
unknown qubit in principle can not be copied\cite{wootters} or 
observed exactly 
therefore the simple repetition code as used in classical coding is not 
applicable here. 
 
With the discovery of maximal polarization entangled state with the
spontaneous parametric down conversion(SPDC)\cite{para},
linear optics method has been perhaps the most powerful tool
for realizing the  entanglement based quantum tasks.
So far many of the tasks have been proposed or demonstrated 
with linear optics, such as quantum teleportation\cite{tele}, 
universal quantum cloning\cite{clone},
quantum U-NOT operation\cite{unot}, quantum entanglement 
concentration and purification\cite{panature,wang0} and destructive 
quantum logic gate\cite{pittman}. However, none of the quantum
error correction code has been 
experimentally realized so far\cite{nmr}. 
Realizing either Shor's 9-qubit code, Steane's 7-qubit code or
the  5-qubit code\cite{puri} is 
technically challenging by our current technology. All of them are based on
the quantum entangled state with more than 5 qubits. 
This requires  at least 3 pairs to be emitted by SPDC\cite{para}.
In a paper two years ago\cite{bbf}, the optical realization of quantum error
rejection code over the bit-flip-error channel is considered.  
It was shown there\cite{bbf} that the controlled-NOT operation
in quantum error correction can be done probabilistically by a polarizing
beam splitter and one can transfer a qubit robustly over a bit flip channel
by teleportation. However, that scheme is based on the resource of three-photon GHZ state which is thought of as a type of impractical resource by our currently existing technology\cite{bbf}. In particular, it was pointed in Ref.\cite{bbf} that the post selection method given by\cite{bghz,pghz} cannot be applied
to the scheme proposed in\cite{bbf}.  In this paper, we propose a realization
of quantum error rejection code over bit-flip-error channel with 
currently existing devices and resources in linear optics.
\section{2-bit bit-flip error rejection code}
To test the main
points of the quantum error correction code  we 
shall consider a simpler case here:  
transmitting an unknown qubit robustly over
the bit flip channel using a smaller quantum  code. 
We assume  no phase flip noise for channel.
Note that even in such a case there is no trivial way to 
complete the task: a repetition code is not allowed by the 
non-cloning principle. 

To further simplify the experimental realization, instead of 
$correcting$ the error, 
here we shall only $reject$ the corrupted qubits
by using an {\it quantum error rejection code} (QERC).      
Suppose Alice is given the following unknown qubit
\begin{eqnarray}\label{u0}
|u\rangle=(\cos(\gamma/2)|0\rangle+e^{i\phi}\sin(\gamma/2)|1\rangle).
\end{eqnarray}
If the qubit is directly sent through the channel, the qubit state
after passing through the bit flip channel will be
\begin{eqnarray}
\rho_a = (1-\eta)|u\rangle\langle u| + \eta |u_f\rangle\langle u_f|
\end{eqnarray}
and 
\begin{eqnarray}\label{uf}
|u_f\rangle =
(\cos(\gamma/2)|1\rangle+e^{i\phi}(\sin\gamma/2)|0\rangle) .
\end{eqnarray}
For all possible initial states on the Bloch sphere,
the average error rate caused by the bit flipping channel is
\begin{eqnarray}
E_0 =1-\frac{1}{\pi}\int_0^{2\pi}\int_{0}^{\pi}
\langle u| \rho_a |u\rangle {\rm d}\gamma {\rm d}\phi
=\frac{2}{3}{\eta}.
\end{eqnarray}

To send the unknown state robustly to the remote party Bob, 
Alice first encodes it
into
\begin{eqnarray}
|q\rangle =(\cos(\gamma/2)|00\rangle+e^{i\phi}\sin(\gamma/2)|11\rangle).
\end{eqnarray}
To make this encoding she does not 
need any information of the given state.
What she needs to do is simply the  conditional unitary 
transformation of 
\begin{eqnarray}
|00\rangle\longrightarrow |00\rangle ;
|10\rangle\longrightarrow |11\rangle,
\end{eqnarray} 
where the first state is the unknown given qubit and the second one is the
ancilla qubit.

She then sends the two-qubit code to the remote party Bob over bit flip
channel, i.e., there is a small probability $\eta$ ($\eta<\frac{1}{2}$) that a qubit is flipped
during the transmission. After Bob receives the code, he first takes a parity 
check on the two qubits: if their bit values are different, he gives up both
of them;
if the bit values are same, he decodes the code by  measuring
the first qubit in code $|q\rangle$ in the basis 
$|\pm\rangle=\frac{1}{\sqrt 2}(|0\rangle \pm |1\rangle)$.
If the result is $|+\rangle$, he stores the second qubit; if the result is 
$|-\rangle$, he takes unitary transformation of 
$(|0\rangle,|1\rangle)\longrightarrow (|0\rangle, -|1\rangle) $  to the 
second qubit and then stores it.
The parity check does not damage the code itself, since the collective measurement only shows whether the two qubits have the same bit value rather
than the bit value information of each qubit.
Note that with the normalization factor being omitted,
\begin{eqnarray}
|q\rangle = |+\rangle|u\rangle + |-\rangle (\cos(\gamma/2)|0\rangle - e^{i\phi}\sin(\gamma/2)|1\rangle).
\end{eqnarray}
 In the case that they have the same bit value, with a relative probability of  $(1-\eta)^2$ that neither quit in the code is flipped, i.e. the code 
state with Bob is still $|q\rangle$. 
With a relative probability in of $\eta^2$ that both of the qubits are flipped,
i.e., the  code state with Bob is
\begin{eqnarray}
|e_1\rangle =(\cos(\gamma/2)|11\rangle+e^{i\phi}\sin(\gamma/2)|00\rangle).
\end{eqnarray}
The cases that one qubit is flipped and one qubit is unchanged will always
lead to different bit values of the two qubits therefore are all discarded
by Bob after the  parity check. 
It can be calculated that  the  average fidelity between the finally 
stored state and the 
initial unknown state is
$
F=\frac{(1-\eta)^2+\eta^2/3}{(1-\eta)^2+\eta^2}.
$
This shows that the error rate after decoding is
\begin{eqnarray}\label{srate}
E_c=\frac{2}{3}\frac{\eta^2}{(1-\eta)^2+\eta^2}.
\end{eqnarray}
However, if Alice directly sends the original qubit without entanglement
based quantum coding,
the error rate will be in the magnitude order of $\eta$, 
which is one order higher than that with quantum rejection code.

Note that the above scheme works for any unknown state including
the case that the initial qubit is entangled with a third party.
\section{Experimental proposal}
We now show the main result of this work: how to experimentally test 
the idea above with practically existing technology
 in linear optics. We propose the quantum
error rejection   scheme in
figure 1. As we are going to show, our scheme works successfully whenever
beam I0, x0 and y0 each contains exactly one photon.  
We are now working in the polarization space, we replace the state notation
$|0\rangle$, $|1\rangle$ by $|H\rangle$, $|V\rangle$ respectively.  
\\{\it 1. Initial state preparation.}
\\When one pair is emitted on each side
of the nonlinear crystal, beam 0,1 and beam 2,3
are both in the entangled state $|\Phi^+\rangle=\frac{1}{\sqrt 2}(|HH\rangle+|VV\rangle)$. 
With the clicking of D0, the initial unknown state 
$\left(\cos(\gamma/2)|H\rangle+e^{i\phi}\sin(\gamma/2)|V\rangle\right)$ 
is prepared
on beam $1'$
\\{\it 2. Encoding.} 
\\After step 1, the state of beam 1',2,3 is 
$\left(\cos\frac{\gamma}{2}|H\rangle+e^{i\phi}\sin\frac{\gamma}{2}|V\rangle\right)
|\Phi\rangle_{23}$. 
The omitted subscripts are $1',2,3$ from left to right to each term.
With the combination of beam $1'$ and beam $2$ by the PBS,
 the state for beam $2',1'',3$ is
\begin{eqnarray}
\frac{1}{\sqrt 2}\left(\cos\frac{\gamma}{2}|H\rangle|H\rangle|H\rangle
+e^{i\phi}|V\rangle\sin\frac{\gamma}{2}|V\rangle|V\rangle
+\cos\frac{\gamma}{2}|0\rangle|HV\rangle|V\rangle
+e^{i\phi}\sin\frac{\gamma}{2}|HV\rangle|0\rangle|H\rangle
\right)\end{eqnarray}. Here the subscripts are implied by 
$|w\rangle|s\rangle|t\rangle=|w\rangle_{2'}|s\rangle_{1''}|t\rangle_3$.
Note that neither vacuum state $|0\rangle_{2'}$ nor two photon
state 
$|HV\rangle_{2'}$ will cause the event of exactly one photon on beam x0.
State $|HV\rangle =a^\dagger_Ha^\dagger_V|0\rangle$.
After a Hadamard transformation by HWP2, beam $2'$ is changed to the  state 
$\frac{1}{\sqrt 2}(|2H\rangle-|2V\rangle)$ on beam I2. This show that beam x0 
contains either 2 photons or nothing, 
if beam $2'$ is in the state $|HV\rangle$.
 Therefore we need only consider the first
two terms above. 
The first two terms above can be rewritten in the equivalent form
of 
\begin{eqnarray}
|+\rangle_{2'}\left(\cos\frac{\gamma}{2}|HH\rangle_{1'',3}+e^{i\phi}\sin\frac{\gamma}{2}|VV\rangle_{1'',3}\right)+
|-\rangle_{2'}\left(\cos\frac{\gamma}{2})|HH\rangle_{1'',3}-e^{i\phi}\sin\frac{\gamma}{2}|VV\rangle_{1'',3}\right). 
\end{eqnarray}  This shows that the state of beam $1'$ is indeed
encoded onto beam $1''$ and beam 3 with the entangled state
$\left(\cos(\gamma/2)|HH\rangle_{1'',3}+e^{i\phi}\sin(\gamma/2)|VV\rangle_{1'',3}|\right)$, 
if beam $2'$ is projected to single photon state $|+\rangle$. 
\\{\it 3. Transmission through the bit flip channel.} 
\\Beam $1''$ and
beam $3$ then each pass through a dashed line rectangular boxes  
which work as  bit flip channels. We shall latter show how the rectangular box
can work as the bit flip channel.
\\{\it 4. Parity check and decoding.} 
\\After the code has passed through the noisy channel, one first take a parity 
check to decide whether to reject it or accept it. To do so one just observe
beam $3''$. If it contains exactly 1 photon, the code is accepted otherwise 
it is rejected. Further, in decoding, one measures beam $3''$ in $|\pm\rangle$
basis (In our set-up this is done by first taking a Hadamard transformation 
to beam $3''$ and then measuring beam I3 in $|H\rangle,|V\rangle$ basis).
If no qubit in the code has been flipped after passing through the channel,the state for beam $1'''$ and beam $3'$ is 
$\left(\cos(\gamma/2)|HH\rangle_{1''',3'}+
e^{i\phi}\sin(\gamma/2)|VV\rangle_{1''',3'}|\right)$ and this state keeps unchanged
after passing through the PBS. Again the state of beam $3''$ and $I1$ can be
rewritten into 
\begin{eqnarray}
|+\rangle_{3''}\left(\cos\frac{\gamma}{2}|H\rangle_{I1}+e^{i\phi}\sin\frac{\gamma}{2}|V\rangle_{I1}\right)+
|-\rangle_{3''}\left(\cos\frac{\gamma}{2}|H\rangle_{I1}-e^{i\phi}\sin\frac{\gamma}{2}|VV\rangle_{I1}\right). 
\end{eqnarray}
If beam $3''$ is projected to state $|+\rangle$, the original state 
is recovered in beam $I1$. Note that if one of the  beam 
in $1'',3$ is flipped, the polarization of beam $1'',3'$ will be either
$H,V$ or $V,H$. This means beam $3''$ will be either in
vacuum state or in the two photon state $|HV\rangle$. Beam I3
will be in the state $\frac{1}{\sqrt 2}(|2H\rangle-|2V\rangle)$ given state
$|HV\rangle$ for beam $3''$. In either case, beam $3''$ or beam y0
 shall never contain exactly 1 photon. This shows that the code
 will be $rejected$  if one qubit
has been flipped. The code with both qubits having been flipped can 
also be accepted, but the probability of 2-flipping is in general very small.
  Therefore the error rate of all
those states decoded from the $accepted$ codes is greatly decreased.
\\{\it 5. Verification of the fault tolerance of QERC}. 
\\To verify the fault tolerance property,
we should observe the error rate of all the accepted qubits. 
The devices Pv-, RPBS, D1 and D4 are  used to measure beam I1 in basis of 
\begin{eqnarray}
|\psi\rangle=\left(\cos(\gamma/2)|H\rangle+e^{i\phi}\sin(\gamma/2)
|V\rangle\right),\nonumber\\
|\psi^\perp\rangle=\left(e^{-i\phi}\sin(\gamma/2)|H\rangle-\cos(\gamma/2)|V\rangle\right).\end{eqnarray}
We shall only check the error rate to the $accepted$ beams. 
For this  we need check whether beam I0,x0 and y0 each contains exactly 
one photon in our scheme. 
The 4-fold clicking  (D0,D2,D3,D1) or (D0,D2,D3,D4) guarantees this.
For simplicity, we shall call the 4-fold clicking  (D0,D2,D3,D1)as event $C_1$
and 4-fold clicking  (D0,D2,D3,D4) as event $C_4$ hereafter.
As we have shown, given the bit flip rate $\eta$, the average error rate
without QERC is $E_0=\eta/2$. The error rate for the accepted qubits with QERC is $E_c$
given by eq.(\ref{srate}). The experimental motivation is to observe the error rate
with our $QERC$  and to demonstrate
this error rate is much less than $E_0$.
The value $E_c$ is obtained by the experiment. 
We shall count the error rate based on the number of each type of four fold events, i.e.,
$C_1$ and $C_4$. Denoting $N_1$, $N_4$ as the observed number of $C_1$ and $C_2$ respectively.
The value $N_4/(N_1+N_4)$ is just the error rate for those accepted qubits with QERC. 
 
The dashed boxes work as   bit flip 
channels. For such a purpose, the phase 
shift $\theta$($-\theta$) or   $\theta_1$($-\theta_1$)
to vertical photon created by P(-P) or P1(-P1)  should be randomly
chosen from $\pm\frac{\pi}{2}$. Note that here a $\Delta$ degree HWP is 
mathematically defined as the unitary 
$U=\left(\begin{array}{cc}\cos\Delta & -\sin\Delta\\
\sin\Delta & \cos\Delta \end{array}\right)$ in the basis of
$\{|H\rangle=\left(\begin{array}{c}1\\0\end{array}\right),|V\rangle=
\left(\begin{array}{c}0\\1\end{array}\right) \}$.
The dashed box  changes the incoming state to 
outgoing state by the following rule:
\begin{eqnarray}
\left(|H\rangle,|V\rangle\right)\longrightarrow \sqrt{\frac{1}{1+\epsilon}}
\left(|H\rangle+\sqrt\epsilon e^{i\theta}|V\rangle, |V\rangle-\sqrt \epsilon e^{-i\theta}|H\rangle\right)
\end{eqnarray}  
Given an arbitrary state 
$|u\rangle= (\cos\frac{\gamma}{2}|H\rangle+e^{i\phi}\sin\frac{\gamma}{2}|V\rangle)$,
 after it passes a dashed square box,
the state is changed to
\begin{eqnarray}
|u_a\rangle = 
\sqrt{\frac{1}{1+\epsilon}}\left[|u\rangle -e^{i\theta}\sqrt {\epsilon} (\cos\frac{\gamma}{2}|V\rangle-
 e^{i\phi}e^{-2i\theta}\sin\frac{\gamma}{2} 
|H\rangle)\right]
\end{eqnarray}
Note that  $e^{-2i\theta}=-1$, since $\theta$ is either 
$\frac{\pi}{2}$ or $-\frac{\pi}{2}$. Since $e^{i\theta}$ takes
the value of $\pm i$ randomly, the state $|u_a\rangle$ is actually in an
 equal probabilistic mixture
of both  $\sqrt{\frac{1}{1+\epsilon}}
\left(|u\rangle \pm i\sqrt \epsilon |u_f\rangle\right)$ therefore the output state of the dashed line square box is 
\begin{eqnarray}
\rho_a=\frac{1}{1+\epsilon}(|u\rangle\langle u|+\epsilon |u_f\rangle\langle u_f|).
\end{eqnarray}
Here $|u_f\rangle $ is defined by Eq.(\ref{uf}) with $|0\rangle, |1\rangle$
being replaced by $|H\rangle,|V\rangle$ respectively.
This shows that the  flipping rate of  the dashed box channel is  
\begin{eqnarray}
\eta=\frac{\epsilon}{1+\epsilon}.
\end{eqnarray} 
Taking average over all possible initial
states on Bloch sphere, the average error rate after a successful decoding 
by our scheme is
\begin{eqnarray}\label{sfidelity}
E_c=\frac{2}{3}\frac{\epsilon^2}{1+\epsilon^2}.
\end{eqnarray}
However, if beam 1' is  directly sent to the remote party through 
one dashed box in our figure,  the average error rate is 
$E_0=\frac{2\epsilon}{3(1+\epsilon)}$, which is much larger than that through
the quantum error rejection code if $\epsilon$ is small.

Although we need a random phase shift of $\pm\frac{\pi}{2}$ for
both $\theta$ and $\theta_1$ in each dashed line square box
to create the 
bit flip channels, in an experiment motivated towards  
detecting the error rate of quantum
error correction code with such a channel, we can simply choose 
$(\theta,\theta_1)=\{(-\frac{\pi}{2},-\frac{\pi}{2}),
(-\frac{\pi}{2},\frac{\pi}{2}),(\frac{\pi}{2},-\frac{\pi}{2}),
(\frac{\pi}{2},\frac{\pi}{2})\}$ separately and run the set-up in each
case for a same duration. 
The average error rate over the total four durations is just the error rate
for  the bit flip channel where $\theta$ is randomly chosen from $\pm\frac{\pi}{2}$. 

The overall efficiency of the experiment
 can be increased by 4 times if we accept all cases of
the initial state preparation and also use the error correction code of
$(\cos(\gamma/2)|HH\rangle - e^{i\phi}\sin(\gamma/2)|VV\rangle)$. To do so we only need to replace 
the polarizer
Ph by a PBS and add one more photon detector there, and also detect
beam x and beam y in the figure.

Our scheme can also be used on the entangled state. To do so we need remove
the devices HWP1 and ${\rm P_{V+}}$, Ph, $\rm {P_{V-}}$, RPBS and D0, D1,D4,
and measure the correlation of beam 0 and beam in 
$\{|H\rangle,|V\rangle\}$ basis and 
$\{|+\rangle,|-\rangle\}$ basis. 
\section{Effects caused by device imperfections}
Now we consider the effects caused by the imperfections including
 limitted efficiency, dark counting of the photon detectors and
 multi-pair (3-pair)  emission in SPDC process. 

The limitted efficiency
of the photon detector only decreases the observable coincidence 
rate but does not
affect the fault tolerance property  of the code. Note that the purpose of
the proposed experiment is to check the error rate to all states which have $passed$
the parity check. This corresponds to the 4-fold coincidence observation. 
If the detecting efficiency is low, many events which $should$ cause the coincidence
would be rejected. That is to say, many good codes will be rejected. But the low detection
efficiency will never cause a corrupted code to pass the parity check. So the net effect of
the low detection efficiency is to reduce the total number of accepted states but it does not
changes the error rate for the accepted qubits. 
In other words, an experiment with limitted detector efficiency is equivalent to  that
with  perfect detectors and a lossy channel.
 Dark counting can be disregarded here because during the coincidence time
 in the order of $10ns$ the dark counting probability is 
less than $10^{-6}$\cite{imoto}. This can always be ignored safely provided
the photon detector efficiency is much larger than $10^{-6}$. Normally, the detector efficiency
is larger than $10\%$, which is much larger than the dark counting rate. 

The probability of 3-pair emission is less than the probability
of 2-pair emission. The probability for
$C_4$ event caused by 2-pair emission is in the magnitude order
of $\epsilon^2 p^2$. The 3-pair emission probability can be comparable
with this if $p$ is not so small.  Also the low detecting efficiency and the encoding-decoding 
process will make 3-pair emission more likely to be observed than 2-pair 
events.  
Now we consider the joint effects of limitted detecting efficiency and 3-pair emission. 
To see the effects, we shall calculate the 
rate of 4-fold events $C_1,C_4$ caused by the 3-pair emission.
Among all 3-pair emissions, 
the cases that all 3 pairs at the same side of the crystal will never cause
the coincidence event.
Three pair states
\begin{eqnarray}
\nonumber
|l\rangle = 
\frac{1}{\sqrt 6}(|H\rangle_0|H\rangle_1+|V\rangle_0|V\rangle_1)
(|2H\rangle_2|2H\rangle_3+|HV\rangle_2|HV\rangle_3+|2V\rangle_2|2V\rangle_3)
\\
|r\rangle = 
\frac{1}{\sqrt 6}
(|2H\rangle_0|2H\rangle_1+|HV\rangle_0|HV\rangle_1+|2V\rangle_0|2V\rangle_1)
(|H\rangle_2|H\rangle_3+|V\rangle_2|V\rangle_3)
\end{eqnarray}
can cause the 4-fold coincidence.
The emission probability of each of them is $3p^3/4$, $p$ is the one-pair 
emission probability. The emission probability for these states are much lower
than that of 2-pair state, $p^2$. However, the 3-pair emission could still 
distort the observed
value of $N_4/(N_1+N_4)$ significantly, since the value $N_4$ itself in the 
ideal case is also small (in the magnitude order of $\epsilon^2p^2$).
We want to verify the fault tolerance property of the error rejection code.
In the ideal case this can be verified by the fact 
that $N_4/(N_1+N_4)<< \eta/2$. To check the joint
effect of 3-pair emission and the limitted detector efficiency, we need simply
to calculate the modification of the rate of 
event $C_4$ by the 3-pair emission and detector efficiency. (Since $N_1$ in the ideal case is
much larger than $N_4$, the 3-pair emission modification to $N_1$ will be disregarded.)
If the modified value of  $N_4/(N_1+N_4)$ is close to the ideal result therefore 
still 
much less than $ E_0=\eta/2$, then we conclude that those imperfections do not
affect the main conclusion of the experiment and the fault tolerance property
of the error-rejection code can be demonstrated even with those imperfections.

Since it will make no difference to the measurement results in average,
for calculation simplicity, 
we postpone all measurements until the code has passed the channels.
And we shall also omit those states which will never cause 4-fold clicking.
Given state $|r\rangle$ or $|l\rangle$ we can write the corresponding 
state on beam  0,I1,$2',3''$. The probability of causing the 4-fold clicking event $C_1$ can then be calculated base on the state of beam
 0,I1,$2',3''$. Note that the state should pass through the bit flip channels
(the dashed rectangular boxes). Therefore given $|r\rangle$ or $|l\rangle$ there could be
4 different state on beam  0,I1,$2',3''$.  Given state $|r\rangle$ initially, 
with probability $(1-\eta)^2$ that  no qubit is flipped
 when passing through the dashed boxes. In such a case, with 
 those terms
 which will never cause 4-fold clicking being omitted, the state
of  beam 0,I1,$2',3''$  will be
\begin{eqnarray}\nonumber
|r\rangle_0=\frac{1}{\sqrt 6}
\{|2H\rangle[\alpha^2|2H,H,H\rangle+\beta^2|V,2V,V\rangle+
\sqrt 2\alpha\beta(|H,HV,H\rangle+|HV,V,V\rangle)]\\
+|HV\rangle[\sqrt 2 \alpha\beta(|2H,H,H\rangle-|V,2V,V\rangle)
+(\beta^2-\alpha^2)(|H,HV,H\rangle+|HV,V,V\rangle)]\}
\label{r0}\end{eqnarray}where
$|\alpha|^2+|\beta|^2=1$ and  for each term we have used the notation 
and subscripts implication as the following:
\begin{eqnarray}
|s\rangle|u,v,w\rangle=|s,u,v,w\rangle
=|s\rangle_0|u\rangle_{I1}|v\rangle_{2'}|w\rangle_{3''}
\end{eqnarray}
In the following, we always imply this order
for the omitted subscripts and omit those components which will never cause 4-fold clicking.
With a probability of $\eta(1-\eta)$ beam $1''$ is flipped, the state is then
\begin{eqnarray}
|r\rangle_{1''}=\frac{1}{\sqrt 6}
\left[\sqrt 2\alpha\beta|2H,HV,V,V\rangle)+
(\beta^2-\alpha^2)|HV,HV,V,V\rangle\right].
\end{eqnarray}
With a probability of $\eta(1-\eta)$ beam $3$ is flipped, the state is then
\begin{eqnarray}
|r\rangle_3 = \frac{1}{\sqrt 6} \left[\sqrt 2 \alpha\beta|2H,H,V,HV\rangle
 +(\beta^2-\alpha^2|HV,H,V,HV\rangle
\right].
\end{eqnarray}
With a probability of $\eta^2$ both beam $1''$ and beam 3 are flipped, the state is then
\begin{eqnarray}\nonumber
|r\rangle_b =\frac{1}{\sqrt 6}|2H\rangle(\alpha^2 |V,H,2V\rangle+ \beta^2|H,2V,H\rangle +\sqrt 2 \alpha \beta |V,HV,V\rangle+\sqrt 2\alpha\beta|H,V,HV\rangle)\\
+\frac{1}{\sqrt 6}|HV\rangle\left[
 \alpha\beta(|V,H,2V\rangle -|H,2V,H\rangle)
+(\beta^2-\alpha^2)(|V,HV,V\rangle+|H,V,HV\rangle)\right].
\end{eqnarray}
Similarly, given initial state $|l\rangle$, we shall also obtain 4 different
states in beam 0,I1,$2',3''$.
If no beam is flipped we have 
\begin{eqnarray}
|l\rangle_0=\frac{1}{\sqrt 6}|H\rangle
\left[\alpha(|H,2H,2H\rangle+|HV,H,HV\rangle) +\beta|V,HV,HV\rangle
+\beta|2V,V,2V\rangle\right]
\end{eqnarray} 
If beam $1''$ is flipped we have
\begin{eqnarray}
|l\rangle_{1''}=\frac{1}{\sqrt 6}|H\rangle
(\alpha|HV,H,HV\rangle+\beta|HV,HV,H\rangle).
\end{eqnarray} 
If beam 3 is flipped we have
\begin{eqnarray}
|l\rangle_3=\frac{1}{\sqrt 6}|H\rangle(\alpha |HV,H,HV\rangle+
\beta|H,HV,HV\rangle).
\end{eqnarray}
If both beam $1''$ and beam 3 are flipped we have
\begin{eqnarray}\label{lb}
|l\rangle_b=\frac{1}{\sqrt 6}|H\rangle\left[\alpha (|2V,2H,V\rangle+|HV,H,HV\rangle)+
\beta(|HV,HV,H\rangle+|2H,V,2H\rangle)\right].
\end{eqnarray}

We have denoted the  4-fold clicking event (D0,D2,D3,D4) by $C_4$.
To calculate the rate of the $C_4$ caused by 3-pair 
states , we just calculate the 4-fold clicking probability caused by each of the above states and then take a 
summation of them.
 Moreover, the probability is
dependent on the parameters in the initial state, $\alpha,\beta$, one should take the average over the whole 
Blosh sphere. 
However, in a real experiment, instead of testing the average over all Bloch sphere, it's
more likely to test the code by the average effect of four state of
\begin{eqnarray}\label{bb84}
(\alpha,\beta)=(1,0); (0,1);\frac{1}{\sqrt 2}(1,1);\frac{1}{\sqrt 2}(1,-1).
\end{eqnarray}
Here we just take the average over these 4 states instead of 
the whole Bloch sphere for simplicity. This obviously will not affect the main
points.

Consider the state $|r\rangle_0$. 
In case of $\alpha=1,\beta=0$, only the state $\frac{1}{\sqrt 6}|2H\rangle|HV,V,V\rangle$ can cause event $C_4$. 
The probability upper bound is
$\frac{1}{24}\xi^4$, where $\xi$ is the detecting efficiency of a photon detector. In the calculation, we have used the fact that 
$|HV\rangle$ is changed to
$\frac{1}{\sqrt 2}(|2H\rangle-|2V\rangle)$ after the Hadamard transformation.
Also, with 2 incident photons, 
a photon detector will be clicked by probability $1-(1-\xi)^2=2\xi-\xi^2$.
To calculate the upper bound, we simply use $2\xi$ and discard $-\xi^2$, this 
will over estimate the clicking probability. However, we 
shall finally show that even with such an overestimation, 
all the $C_4$ events caused by 3-pair emission will not affect the main 
results. Similarly, if $\beta=1,\alpha=0$, the probability to cause $C_4$
events by state $|r\rangle_0$ is also upper bounded by
$\frac{1}{24}\xi^4$. Now we consider the case of $\alpha=\beta=\frac{1}{\sqrt 2}$.
After calculation
we find that the probability of $C_4$ event caused by each term in $|r\rangle_0$ is upper bounded by the following table
the 4-fold clicking. The probability to cause the 4-fold clicking by each term is listed in the following table:
\begin{center}
\begin{tabular}{rccccccccc}
term & 1 & 2 & 3 & 4 & 5 & 6 & 7 & 8 & Sum.
\\ 
\hline 
  Prob. & $\frac{\xi^2}{48}$ & $\frac{\xi^4}{48}$ & $\frac{\xi^4}{24}$ & $\frac{\xi^4}{24}$ & $\frac{\xi^4}{96}$ & $\frac{\xi^4}{96}$ & 0 &0 
& $\frac{7\xi^4}{48}$
\end{tabular}
\end{center} 
 Similarly, in the case of $\alpha=-\beta=\frac{1}{\sqrt 2}$ the probability to cause $C_4$ event is also $7\xi^4/48$.
Therefore in average the probability of $C_4$ events caused by state $|r\rangle_0$ is $\frac{3\xi^4}{16}$.
We list the upper bound of 
average probability contribution to $C_4$ events caused by each state 
 from eq.(\ref{r0}) to eq.(\ref{lb})
in the following
\begin{center}
\begin{tabular}{rcccccccc}
state & $|r\rangle_0$ & $|r\rangle_{1''}$ & $|r\rangle_3$ & $|r\rangle_b$ & $|l\rangle_0$ & $|l\rangle_{2''}$ & $|l\rangle_3$ & $|l\rangle_b$ 
\\ 
\hline 
  Prob. & $\frac{3\xi^4}{16}$ & $\frac{\xi^4}{16}$ & $\frac{\xi^4}{16}$ & $\frac{17\xi^4}{96}$ & 
$\frac{5\xi^4}{24}$ & $\frac{\xi^4}{12}$ & $\frac{\xi^4}{12}$ & $\frac{5\xi^4}{24}$ 
\end{tabular}
\end{center} 

Suppose the detecting efficiency of the photon detector is $\xi$.
With two incident photons, the photon detector will be clicked with probability $R_2=2\xi-\xi^2<2\xi$.
For calculation simplicity we shall use $2\xi$ to replace $R_2$. This will overestimate the effect caused by
 3-pair emission.  Also we shall  count all 5-fold clicking events as $C_4$, this will further overestimate
the 3-pair emission effect because in a real experiment one may discard all 5-fold clicking events.
With these two overestimation, what we shall calculate is the upper bound of the detectable error rate
with 3-pair emission and limited detector efficiency being taken into consideration. 
The total probability of $C_4$ event caused by all 3-pair emission is upper bounded by
\begin{eqnarray}
\lambda_3=\xi^4\left[ (1-\eta)^2 \cdot\frac{19}{48} +(1-\eta)\eta\cdot\frac{7}{24}
+\eta^2\cdot\frac{17}{48}\right]
\cdot \frac{3p^3}{4}.
\end{eqnarray}
We have known that the probability of $C_4$ events caused 2-pair emission is $\lambda_2=\frac{1}{16}\cdot \xi^4\eta^2p^2$, which corresponds
 to the error rate in the idea case, i.e. eq.(\ref{sfidelity},\ref{srate}).
Therefore the observed value for $N_4/(N_1+N_4)$ will be upper bounded by 
\begin{eqnarray}
E_c'=(1+\lambda_3/\lambda_2)E_c.
\end{eqnarray}
Note that for whatever detection efficiency, the observed error rate is always upper bounded by $E_c'$.
The $observed$ error rate is  higher
than that in the ideal case due to the joint effect of non-perfect detection efficiency and the 3-pair emission. 
However, this   does not affect
the main result in a real experiment. As one may see from Fig.(2), with  the QERC, 
the observed upper bound of error rate $E_c'$ is very close to the ideal one if the one pair emission rate is not larger
than $0.002$. Given such an emission rate, one may collect dozens of 4-fold clicking data per hour.
\section{concluding remarks}  
In summary, we have shown how to encode and decode a type of
2-qubit quantum error rejection code with spontaneous parametric down 
conversion. 
In our scheme, we require beam $3''$ and beam $I1$ each contain exactly
1 photon. To verify this by our current technology we have no choice but 
to detect both of them. This means  that 
the result is tested by post selection.
However, as it was pointed out in Ref.\cite{bbf}, 
even a post selection result here has a wide application background
such as the quantum cryptography and quantum communication. The details
of the application of the post-selection quantum error rejection code 
in quantum cryptography with hostile channel
has been studied in \cite{bbf}. Obviously, if our scheme is used for
quantum key distribution (QKD), the threshold of error rate\cite{gl} of noisy channel
is improved. A modified scheme can be used to reject the phase flip
error. This may help to improve the tolerable channel flip rates
of Gottesman-Lo protocol\cite{gl}. Details of this have been 
reported elsewhere\cite{wang12}. Note that for the purpose of QKD,
the encoding process can be omitted. One directly produces and sends
the 2-bit code. In such a way, we may transmits thousands of 2-bit codes
per second by our currently existing technology. \\
{\bf Acknowledgement:} We thank Prof. H. Imai  for support.
I thank Dr. B.S. Shi for pointing out Ref.\cite{bbf}. 
I thank Dr H Fan, Dr K. Matsumoto and Dr A. Tomita for discussions. 

\begin{figure}
\begin{center}
\epsffile{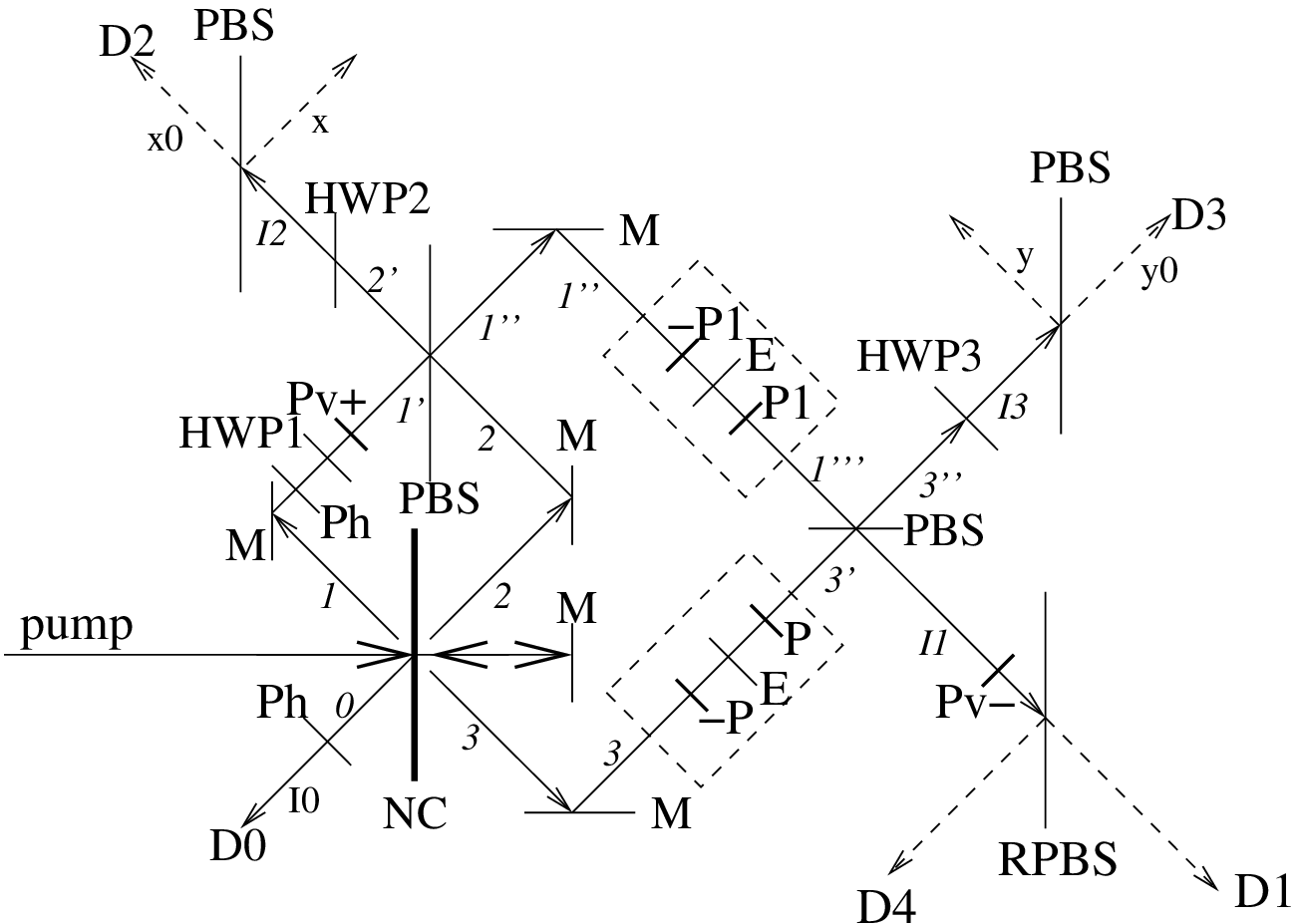}
\end{center}
\caption{Realizing QERC with SPDC process. 
 If beams I0, x0 and y0 each contain exactly one photon, beam I1 is 
accepted, otherwise it is rejected.  The error rate of all the accepted
beams is the $N_4/(N_1+N_4)$, where $N_1$ and $N_4$ are the number of 4-fold clicking
of (D0,D2,D3,D1)  and  (D0,D2,D3,D1) respectively. The dashed rectangular boxes play the role of bit flip channels.
 NC: nonlinear crystal used in  SPDC process. 
M: mirror. Ph: horizontal polarizer.
 HWP2 and HWP3: $\pi/4$ half
wave plates.
HWP1:  $\gamma/2$ half wave plate. 
Pv+,Pv-: $\phi,-\phi$ phase shifters  
to vertically polarized photons only.
PBS: polarizing beam splitter which transmits the horizontally polarized
photons and reflects the vertically polarized photons. D0,D1,D2,D3,D4: photon detectors.
 RPBS: rotated polarizing beam splitter which transmits
the photon in the state 
$\cos\frac{\gamma}{2}|H\rangle + \sin\frac{\gamma}{2} |V\rangle)$ and reflects
the photon in state $ (\sin\frac{\gamma}{2} |H\rangle - \cos\frac{\gamma}{2}|V\rangle)$.
 P, -P, P1 and -P1: phase
shifters, each of 
 takes a   phase shift $\theta,-\theta, \theta_1,-\theta_1$ respectively
to a vertically polarized 
photon only.
 $\theta,\theta_1$ each is a random value from $\pm\frac{\pi}{2}$ .
  E: $\sin^{-1}\frac{\sqrt \epsilon}{\sqrt {1+\epsilon}}$
 half wave plate.  }
\end{figure}
\begin{figure}
\begin{center}
\epsffile{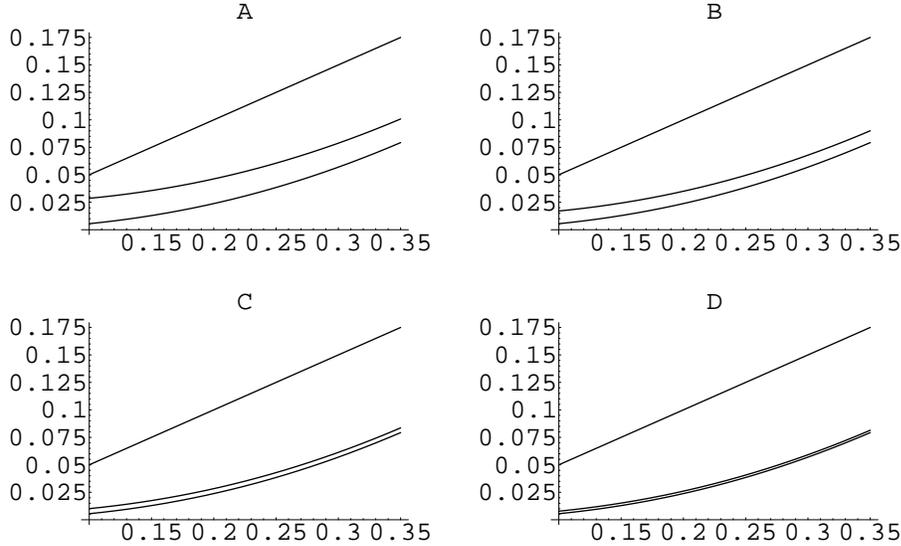}
\end{center}
\caption{Effects of 3-pair emission and limitted detection efficiency to  practical
 experiments. The horizontal  axis is for the bit flipping rate $\eta$ of the channel.
The vertical axis is for the  error rates. The top straight line is for $E_0$: the expected result in the
case that all qubits are sent directly through the bit flip channel, 
without using QERC. The lowest curve is for $E_c$: the expected result in the idea case: sending the qubit with
perfect QERC. 
The curve upper to the lowest curve is for $E_c'$, the upper bound of the expected result
in the practical case of sending qubits with a non-perfect QERC with SPDC process. 
All calculations are done by taking average over the 4 states of  eq.(29).
The distortation
comes from the 3-pair emission and the limitted efficiency of the photon detectors.
Fig. A,B,C,D are for the case of one pair emission probability $p=1/100, 5/1000, 2/1000,1/1000$ respectively in the SPDC process. 
Note that with whatever detection efficiency, the observed error rate is always upper bounded by $E_c'$.
We can obviously see that the distortion caused by  3-pair emission and the limitted detecting efficiency are negligible
when $p\le 2/1000$. }
\end{figure}

\begin{thebibliography}{99}
\bibitem{nmr} The NMR demonstration is not a strict demonstration since there
is no quantum entanglement involved.
\end{thebibliography}
\end{document}